\documentstyle[epsf]{elsart}

\begin{document}
\begin{frontmatter}
\title{Anomalous scaling in a shell model of helical turbulence}

\author{P. D. Ditlevsen\thanksref{mail1}}

\address{Niels Bohr Institute, Geophysical Department, Juliane Maries Vej 30, 2100 Copenhagen O, Denmark}

\author{P. Giuliani}
\address{Dipartimento di Fisica and Istituto Nazionale di Fisica della Materia, 
Universit\`{a} della Calabria, 87036 Rende (CS), 
Italy}

\thanks[mail1]{corresponding author, pditlev@gfy.ku.dk}

\begin{abstract}
In a helical flow there is a subrange of the inertial range in which
there is a cascade of both energy and helicity. In this range the
scaling exponents associated with the cascade of helicity can be
defined. These scaling exponents are calculated from a simulation
of the GOY shell model. The scaling exponents for even moments are
associated with the scaling of the symmetric part of the probability
density functions while the odd moments are associated with the 
anti-symmetric part of the probability density functions.

\end{abstract}


\end{frontmatter}

\section{Introduction}
In a helical flow both energy and helicity are inviscid invariants which are
cascaded from the integral scale to the dissipation scale \cite{Lesieur}. If these scales for
the helicity are separated there will be an inertial range in which an 
equivalent of the four-fifth law for helicity transfer holds. This is a 
scaling relation for a third order structure function with a different
tensorial structure from the structure function associated with the flux of energy.
For helicity flux this is, 
$\langle \delta {\bf v}_\|(l)\cdot[{\bf v}_\bot(r)\times {\bf v}_\bot(r+l)]\rangle 
= (2/15) \overline{\delta}l^2$, where $\overline{\delta}$ is the mean dissipation
of helicity. This relation is called the 'two-fifteenth
law' due to the numerical prefactor \cite{Procaccia,russian}. 
The inertial ranges for helicity cascade and for energy cascade are different
because the dissipation of helicity scales as 
$D_H(k) \sim k D_E(k)$, thus the helicity will be dissipated within
the inertial range for energy cascade. From balancing the helicity flux
and the helicity dissipation a Kolmogorov scale $\xi=K_H^{-1}$ for helicity dissipation
can be defined \cite{D&G},

\begin{equation}
\xi \sim (\nu^3 \overline{\varepsilon}^2/\overline{\delta}^3)^{1/7},
\label{KH}
\end{equation}
where $\nu$ is the kinematic viscosity, $\overline{\varepsilon}$ is
the mean energy dissipation per unit mass and $\overline{\delta}$ is the mean
helicity dissipation per unit mass.
This scale is larger than the usual Kolmogorov scale 
$\eta=K_E^{-1}\sim (\nu^3/\overline{\varepsilon})^{1/4}$.

The physical
picture for fully developed helical turbulence is shown schematically in
figure 1. The mean dissipations 
$\overline{\delta}$ and $\overline{\varepsilon}$ are solely determined by
the forcing in the integral scale. There will then be an inertial range with
coexisting cascades of energy and helicity with third order structure functions
determined by the four-fifth -- and the two-fifteenth laws. This is followed
by an inertial range between $K_H$ and $K_E$ corresponding to non-helical turbulence,
where the dissipation of positive and negative helicity vortices balance and
the two-fifteenth law is not applicable.

%\begin{figure}
%\vspace {7 truecm}
%% \hskip 1truecm\special {ps: /fin2/fig21.ps x=5 truecm y=5truecm}
%\special {ps: paladin1.eps x=5 truecm y=5truecm}
%\caption
%{The inertial range for helicity cascade is smaller than the inertial
%range for energy cascade. In the range $K_H<k<K_E$ the dissipation
%of positive and negative helicity balance.}
%\bigskip
%\end{figure}

\begin{figure}[htb]
\epsfxsize=13.5cm
\epsffile{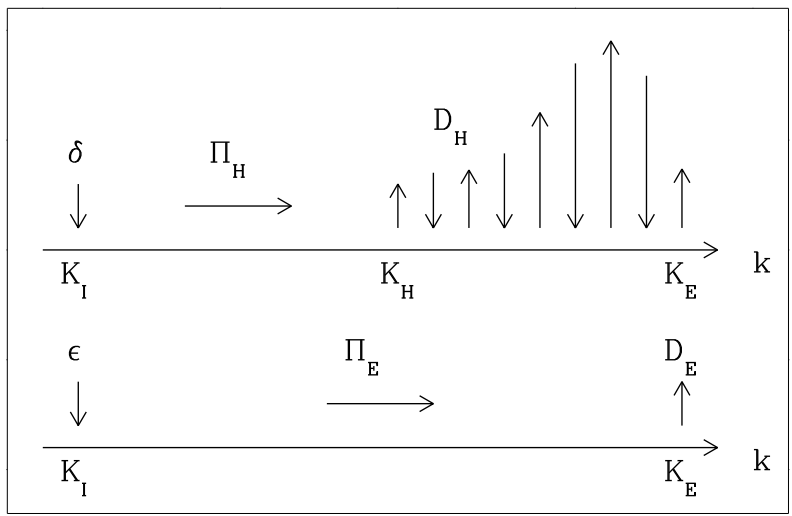}
\caption[]{The inertial range for helicity cascade is smaller than the inertial
range for energy cascade. In the range $K_H<k<K_E$ the dissipation
of positive and negative helicity balance.}
\end{figure}

\section{The anomalous scaling exponents}

There is now experimental evidence that the K41 scaling 
relations are not exact. There are corrections for moments different
from 3, expressed through anomalous scaling exponents, $\langle \delta
v(l)^p \rangle \sim l^{\zeta(p)}$ where $\zeta(p)\ne p/3$. 
Understanding and quantitatively determining the anomalous scaling exponents is
one of the most intriguing and unsolved problems in turbulence.
The intermittency corrections to the K41 scaling could depend on the transfer of
helicity, maybe similar to the way the different
sectors in anisotropic turbulence might give rise to sub-leading corrections
of scaling exponents \cite{Arad}. 
Furthermore,
the helicity cascade 
itself leads to a set of anomalous scaling exponents related to moments 
of the third order correlator of the two-fifteenth law. 
There is at present no experimental measurements from helical turbulence of the
scaling exponents associated with the two-fifteenth law. 

It was recently shown numerically 
by Biferale et al. \cite{Biferale} that in the case of a shell model the anomalous
scaling exponents for the helicity transfer has a strong difference
between odd and even powers such that the scaling exponent $\zeta^H(p)$ is not a convex 
function. 

Biferale et al. used a shell model consisting of two coupled GOY shell models. We
will show here that the results obtained holds for the standard GOY shell
model as well. 
Shell models are
toy-models of turbulence which by construction have second order inviscid
invariants similar to energy and helicity in 3D turbulence.
Shell models
can be investigated numerically for high Reynolds numbers,
in contrast to the Navier-Stokes equation, so that high order statistics
and anomalous scaling exponents are easily accessible.
Shell models lack any spatial structures so
we stress that only certain aspects of the turbulent cascades have
meaningful analogies in the shell models. This should especially
be kept in mind when studying helicity which is intimately linked
to spatial structures, and the dissipation of helicity to reconnection of
vortex tubes \cite{Levich}. The following thus only concerns the spectral aspects
of the helicity and energy cascades.

The GOY model \cite{GOY,Kadanoff,jpv} is the most well studied shell model.
It is defined from the governing equation,

\begin{equation}
\dot{u_n}=i k_n (u_{n+2}u_{n+1}-\frac{\epsilon}{\lambda}u_{n+1}u_{n-1}+
\frac{\epsilon-1}{\lambda^2}u_{n-1}u_{n-2})^* -\nu k_n^2 u_n + f_n
\label{1}
\end{equation}
with $n=1, ..., N$ where the $u_n$'s are the complex shell velocities. The
wave numbers are defined as $k_n = \lambda^n$, where $\lambda$ is the shell
spacing. The second and third terms are dissipation and forcing. The model
has two inviscid invariants, energy, $E=\sum_n E_n =\sum_n |u_n|^2$,
and 'helicity', $H=\sum_nH_n=\sum_n (\epsilon -1)^{-n}|u_n|^2$. The
model has two free parameters, $\lambda$ and $\epsilon$. The 'helicity'
only has the correct dimension of helicity if $|\epsilon -1|^{-n}=k_n
\Rightarrow 1/(1-\epsilon)=\lambda$.
In
this work we use the standard parameters $(\epsilon,\lambda)=(1/2,2)$
for the GOY model.

A natural way to define the structure functions of moment $p$ is through the
transfer rates of the inviscid invariants,

\begin{eqnarray}
S^E_p(k_n)= \langle(\Pi^E_n)^{p/3}\rangle k_n^{-p/3}\sim k_n^{-\zeta^E(p)} \label{s3e} \\
S^H_p(k_n)= \langle(\Pi^H_n)^{p/3}\rangle k_n^{-2p/3}\sim k_n^{-\zeta^H(p)}
\label{s3h}
\end{eqnarray}
The energy flux is defined in the usual
way as $\Pi^E_n = d/dt|_{n.l.}(\sum_{m=1}^{n} E_m) $ where
$d/dt|_{n.l.}$ is the time rate of change due to the non-linear term in
(\ref{1}). The helicity flux $\Pi^H_n$ is defined similarly. By a simple algebra we have
the following expression for the fluxes,

\begin{eqnarray}
\langle \Pi^E_n \rangle= (1-\epsilon) \Delta_n + \Delta_{n+1} =\overline{\varepsilon}\label{pie} \\
\langle \Pi^H_n \rangle= (-1)^n k_n(\Delta_{n+1}-\Delta_n)=\overline{\delta}
\label{pih}
\end{eqnarray}
where $\Delta_n = k_{n-1}Im \langle u_{n-1}u_nu_{n+1}\rangle$,
$\overline{\varepsilon}$ and $\overline{\delta}$ are the mean dissipations of
energy and helicity respectively. The first equalities hold without averaging
as well.
These equations are the shell model equivalents of
the four-fifth -- and the two-fifteenth law. 

In the definition (\ref{s3e}), (\ref{s3h}) of the structure functions there is a slight
ambiguity in the definition of
$x^{p/3}$ for negative $x$ and $p$ not a multiplum of 3.
The complex  roots for
$(-1)^{1/3}$ are $(-1, 1/2 \pm i \sqrt{3}/2)$ and for $(-1)^{2/3}$ they are $(1,- 1/2 \pm i \sqrt{3}/2)$. The common way of circumventing the ambiguity
is by defining
$x^{p/3}= sgn(x)|x|^{p/3}$, which neglects the imaginary 
roots\footnote{This can not
always be done. Had we defined the structure functions from some, say, 
sixth order correlator, we are in trouble since
$z=(-1)^{1/6}$ has no real roots.}. 
With this
definition we have,

\begin{eqnarray}
S_p(k_n) = \int_0^{\infty}[\psi_n(x)+\psi_n(-x)]x^{p/3}dx \equiv \int_0^{\infty}\psi_n^+(x)x^{p/3}dx \, \,
&(p\, \mbox{ even}) \label{pdf1} \\
S_p(k_n) = \int_0^{\infty}[\psi_n(x)-\psi_n(-x)]x^{p/3}dx \equiv 
\int_0^{\infty}\psi_n^-(x) x^{p/3}dx \, \,
&(p\, \mbox{ odd}) \label{pdf2}
\end{eqnarray}
where $\psi_n(x)$ is the probability density function (pdf) for $\Pi_n$.
$\psi_n^+(x)$ is (twice) the symmetric part of the pdf and $\psi_n^-(x)$
is (twice) the anti-symmetric part. Note that $\psi_n^+(x)$ is itself
a pdf while $\psi_n^-(x)$ is not. $\psi_n^-(x)$ is, except for a normalization, a pdf only if $\psi_n(x)>\psi_n(-x)$ for
all positive $x$. 

\begin{figure}[htb]
\epsfxsize=13.5cm
\epsffile{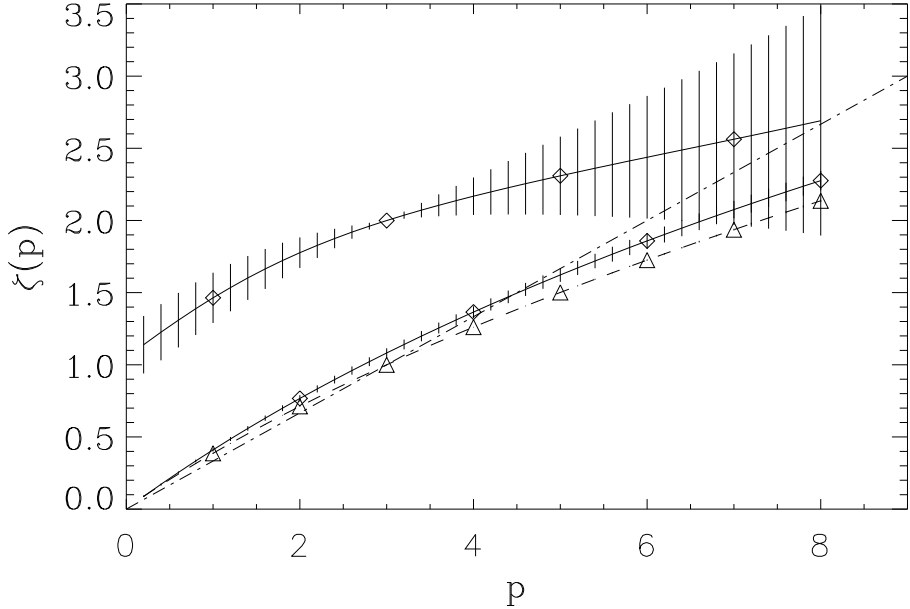}
\caption[]{The anomalous scaling exponents, $\zeta^E(p)$ (dashed
curve) and $\zeta^H(p)$. The two full curves for even and odd moments
for the helicity are calculated according to (\ref{pdf1}) and (\ref{pdf2})
respectively. The error bars on the $\zeta^E(p)$ curve are
within the triangles, so we see a deviation between
$\zeta^E(2p)$ and $\zeta^H(2p)$.
The error bars for the odd moments of $\zeta^H(p)$ are
large because they are determined by the scaling of the anti-symmetric part of the probability
density functions. The dashed-dotted line is the K41 scaling.}
\end{figure}

The scaling exponents are determined from the scaling of the pdf's through,

\begin{equation}
\int_{-\infty}^{\infty}x^{p/3}\psi_{\lambda k}(x)dx=\lambda^{-\zeta(p)}
\int_{-\infty}^{\infty}x^{p/3}\psi_k(x)dx,
\end{equation}
so the scaling exponents for $p$ even is related to the scaling of $\psi_n^+$
while for $p$ odd they are related to the scaling of $\psi_n^-$.
We have performed a simulation of the standard GOY model with 
$(\epsilon, \lambda, \nu, N)=(1/2, 2, 10^{-9}, 26)$ and a forcing
of the form $f_n = 0.01 \delta_{2,n}/u_2^*$, corresponding to a constant 
energy input. The simulation was about $5000$ large eddy turnover times.

Figure 2 shows the anomalous scaling exponents, $\zeta^E(p), \zeta^H(p)$
for the energy and the
helicity calculated according to (\ref{pdf1}) and (\ref{pdf2}). 
Using (\ref{pdf1}) the
scaling exponent $\zeta^H(p)$ can be defined for any real positive $p$, which
from the H\"{o}lder inequality is a convex curve. Similarly using (\ref{pdf2})
assuming $\phi_n^-(x)$ to be a positive function we can define a continuous 
curve $\tilde{\zeta}^H(p)$ which is also from the H\"{o}lder inequality 
a convex curve. 
The scaling exponent $\zeta^H(p)$ defined for integer $p$ jumps between the two curves shown in figure 2.
The scaling exponents differs from the ones found by
Biferale et al. for the two-component GOY model. We find that
$\zeta^H(2p)$ is slightly larger than $\zeta^E(2p)$. 
The scaling regime in
which $\zeta^H(p)$ is calculated is $K_{\mbox{I}}<k<K_H$ while for the
energy $K_{\mbox{I}}<k<K_E$.
The negative part of the probability density is negligible in the
case of energy transfer, $\psi_n(x)\approx \psi_n^+(x)\approx \psi_n^-(x)$ for
$x>0$, but for helicity transfer the negative tail is big which gives
the strong even-odd oscillations between the two curves. Note that $\zeta^E(3)=1$ and $\zeta^H(3)=2$
are just the four-fifth -- and the two-fifteenth law.

\begin{figure}[htb]
\epsfxsize=13.5cm
\epsffile{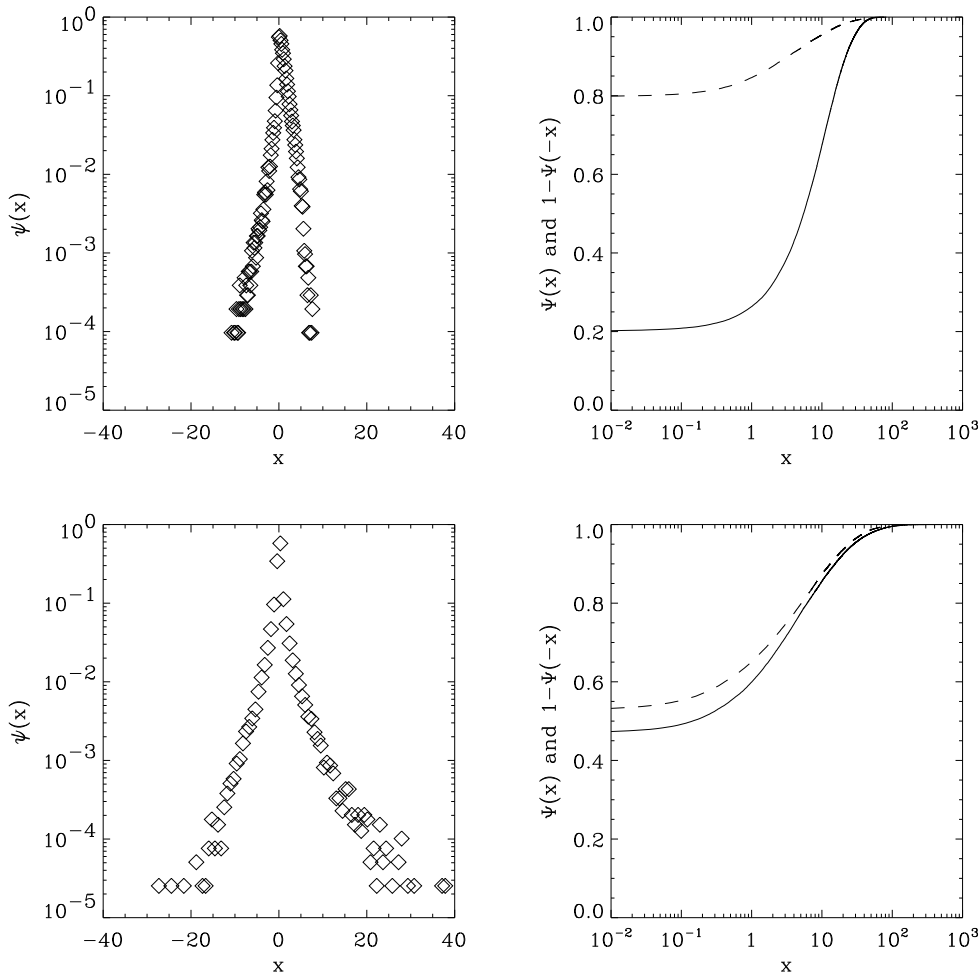}
\caption[]{The left hand panels show probability density 
functions $\psi(x)$ of the
helicity flux for shells 3 and 6. The
right hand panels show the probability distribution functions $\Psi(x)$ for $x > 0$,
the dashed curves are $1-\Psi(-x)$ which for a symmetric distribution is identical
to $\Psi(x)$. The scaling exponents for the even moments depends on the scaling
with $n$ of the symmetric part of the pdf which corresponds to the mean curve between the full and
dashed curves. The scaling exponents for the odd moments depends on the scaling of
the anti-symmetric part of the pdf corresponding to the gap betwwen the full and dashed curves.
}
\end{figure}

Figure 3 shows the probability distribution function (PDF) for helicity flux, 
defined by $\Psi(x)=\int_{-\infty}^x \psi(y)dy$ for shell numbers $n=3$ and
$n=6$ both
in the inertial range for helicity flux.
The negative tail is plotted as $1-\Psi(-x)$ which for a symmetric pdf gives 
two overlapping curves. We can similarly define the PDF $\Psi_n^\pm(x)\equiv 
\int_0^x\psi_n^\pm(y)dy$. A simple algebra gives $\Psi_n^+(x)=\Psi_n(x)+(1-\Psi_n(-x))-1$
and $\Psi_n^-(x)=\Psi_n(x)-(1-\Psi_n(-x))+(1-2\Psi(0))$. So we see that
the scaling of $\Psi_n^+(x)$ is related to the scaling of the mean of the
two curves in the right panels in figure 3, while the scaling of $\Psi_n^-(x)$
is related to the gap between the two curves. 

\section{Conclusion}
Coexisting cascades of energy and helicity are possible in the GOY
shell model. The scaling of the odd order moments of the helicity
transfer depends on the scaling of the anti-symmetric part of the probability
density function for the helicity flux. This defines a convex 
anomalous scaling curve through the point $\zeta^H(3)=2$ which is
the two-fifteenth law. The even order moments of the helicity
flux has anomalous scaling exponents
close to the ones found for the energy flux. In the simulation
a scale break at $K_H$ is not observed. This implies that the anomalous
scaling exponents for the energy flux are not influenced
by the cascade of helicity.

%\begin{ack}

%\end{ack}


\begin{thebibliography}{99}

\bibitem{Lesieur}M. Lesieur, 'Turbulence in Fluids', Third edition, Kluwer Academic Publishers, 1997.
\bibitem{Procaccia}V. S. L'vov
et al. chao-dyn/9705016 (unpublished).
\bibitem{russian}O. Chkhetiani, JETP Lett., 63, 808, 1996.
\bibitem{D&G}P. D. Ditlevsen and P. Giuliani, chao-dyn/9910013.

%\bibitem{Frisch}U. Frisch, 'Turbulence, The Legacy of A. N. Kolmogorov',
%Cambridge Univ. Press., 1995.
\bibitem{Arad}I. Arad et al., Phys. Rev. Lett., 82, 5040, 1999.
\bibitem{Biferale}L. Biferale, D. Pierotti, and F. Toschi,
J. Phys. IV France, 8, Pr6-131, 1998.
\bibitem{Levich}E. Levich, L. Shtilman and A.V. Tur, Physica A, 176, 241, 1991.
%\bibitem{PD}P. D. Ditlevsen, Phys. Fluids, 9, 1482, 1997.
\bibitem{GOY}E. B. Gledzer, Sov. Phys. Dokl, 18, 216, 1973.
\bibitem{Kadanoff}L. Kadanoff et al., Phys. Fluids, 7, 617, 1995.
\bibitem{jpv}M. H. Jensen, G. Paladin and A. Vulpiani, Phys. Rev. A, 43, 798, 1991.
%\bibitem{Biferale}L. Biferale, D. Pierotti, and F. Toschi, Phys. Rev. E, 57, R2515, 1998.
%\bibitem{Olla}P. Olla, Phys. Rev. E, 57, 2824, 1998.

\end{thebibliography}
\end{document}